\documentclass[aps,prd,showpacs]{revtex4}
\usepackage{epsf,epsfig,graphicx}
\begin{document}
\preprint{CU-Phys/05/2007}
\title{Next-to-leading-order corrections to exclusive processes in
$k_T$ factorization}

\author{Soumitra Nandi$^{1}$ and Hsiang-nan Li$^{2}$}
\email{hnli@phys.sinica.edu.tw} \affiliation{$^{1}$Department of
Physics, University of Calcutta, 92 A.P.C Road, Kolkata 700009,
India} \affiliation{$^2$Institute of Physics, Academia Sinica,
Taipei, Taiwan 115, Republic of China,} \affiliation{Department of
Physics, National Cheng-Kung University, Tainan, Taiwan 701,
Republic of China} \affiliation{Department of Physics, National
Tsing-Hua University, Hsinchu, Taiwan 300, Republic of China}

\begin{abstract}

We calculate next-to-leading-order (NLO) corrections to exclusive
processes in $k_T$ factorization theorem, taking
$\pi\gamma^*\to\gamma$ as an example. Partons off-shell by $k_T^2$
are considered in both the quark diagrams from full QCD and the
effective diagrams for the pion wave function. The gauge
dependences in the above two sets of diagrams cancel, when
deriving the $k_T$-dependent hard kernel as their difference. The
gauge invariance of the hard kernel is then proven to all orders
by induction. The light-cone singularities in the $k_T$-dependent
pion wave function are regularized by rotating the Wilson lines
away from the light cone. This regularization introduces a
factorization-scheme dependence into the hard kernel, which can be
minimized in the standard way. Both the large double logarithms
$\ln^2k_T$ and $\ln^2 x$, $x$ being a parton momentum fraction,
arise from the loop correction to the virtual photon vertex, the
former being absorbed into the pion wave function and organized by
the $k_T$ resummation, and the latter absorbed into a jet function
and organized by the threshold resummation. The NLO corrections
are found to be only few-percent for $\pi\gamma^*\to\gamma$, if
setting the factorization scale to the momentum transfer from the
virtual photon.

\end{abstract}


\maketitle

\section{INTRODUCTION}

$k_T$ factorization theorem \cite{CCH,CE,LRS,BS,LS,HS}, as a
fundamental tool of perturbative QCD (PQCD), has been widely
applied to inclusive and exclusive processes. It has been pointed
out that $k_T$ factorization theorem is appropriate for processes
dominated by contributions from small parton momentum factions $x$
\cite{NL2}. Its application to exclusive $B$ meson decays has led
to the PQCD approach \cite{LY1,CL,YL,KLS,LUY}, which is free of
the singularities from the end-point regions of $x$ that usually
appear in collinear factorization theorem
\cite{BL,ER,CZS,CZ,BBNS,BPS}. Several aspects of $k_T$
factorization theorem have been studied. For example, a naive
definition of $k_T$-dependent hadron wave functions, in which the
coordinate of a quark field is simply shifted by a transverse
distance, contains light-cone divergences \cite{Co03}. Modified
definitions to remove these divergences have been proposed in
\cite{Co03,LL04,MW}. The $B$ meson wave function defined in $k_T$
factorization theorem is normalizable \cite{LL04}, while the $B$
meson distribution amplitude in collinear factorization theorem is
not \cite{Neu03,BIK}, when evolution effects are taken into
account. The Sudakov resummation \cite{BS,LY1,CS,MR} of the large
double logarithm $\ln^2 k_T$ is essential for improving
perturbative expansion in $k_T$ factorization theorem
\cite{TLS,WY}.

The current application of $k_T$ factorization theorem to
exclusive processes is mainly made at leading order (LO) in the
strong coupling constant $\alpha_s$ \cite{LRev}: the important
logarithms in hadron wave functions have been organized to all
orders, but hard kernels are still evaluated at tree level. To
demonstrate that $k_T$ factorization theorem is a systematical
tool, higher-order calculations of hard kernels are demanded. In
this paper we shall elucidate the framework for these
calculations, deriving the next-to-leading-order (NLO) hard kernel
for the scattering process $\pi\gamma^*\to\gamma$ as an example.
The point is that partons in both the quark diagrams from full QCD
and the effective diagrams for the pion wave function, carrying
the momentum $k=(k^+,0, {\bf k}_T)$, are off mass shell by
$k_T^2$. The difference between the two sets of diagrams defines
the hard kernel in $k_T$ factorization theorem, a procedure
similar to the derivation of Wilson coefficients in an effective
field theory. This is the way to obtain a $k_T$-dependent hard
kernel without breaking gauge invariance, since the gauge
dependences cancel between the above two sets of diagrams. A
physical quantity is expressed as a convolution of a hard kernel
with model wave functions, which are determined by methods beyond
a perturbation theory, such as lattice QCD and QCD sum rules, or
extracted from experimental data. A gauge-invariant hard kernel
then leads to gauge-invariant predictions from $k_T$ factorization
theorem.

We emphasize that the above prescription for computing a
$k_T$-dependent gauge-invariant hard kernel has not yet been fully
recognized. Several NLO calculations, which include the transverse
momentum dependence via on-shell partons carrying $k=(k^+,k^-,
{\bf k}_T)$, $k^-=k_T^2/(2k^+)$, have been performed in the
literature \cite{MW,KPY,MW0607}. In these calculations both quark
diagrams and effective diagrams are gauge-invariant, and so are
hard kernels. However, the considered parton momentum is not a
configuration described by the nonlocal matrix elements associated
with $k_T$-dependent hadron wave functions, because the minus
component $k^-$ should have been integrated out. Another subtlety
is that the NLO hard kernel for the process $\pi\gamma^*\to\gamma$
obtained in the above formalism turns out to be $k_T$-independent
\cite{MW0607}. The parton transverse degrees of freedom in the
pion wave function are then integrated out, and the formalism
reduces to collinear factorization theorem. Moreover, we shall
explain that the additional nonperturbative soft function
introduced in \cite{MW0607} is not necessary for $k_T$
factorization theorem, since the infrared logarithms can be
absorbed into the pion wave function completely.

As stated before, the light-cone singularities \cite{Co03} in the
naive definition for $k_T$-dependent hadron wave functions must be
regularized. These singularities, not present in the quark
diagrams, are not physical. If not regularized, higher-order hard
kernels, computed as the difference of the quark diagrams and the
effective diagrams, will be divergent. In this paper we shall
adopt the modified definition, in which the Wilson lines involved
in the nonlocal matrix elements for hadron wave functions are
rotated away from the light cone. After the subtraction of the
singularities, a hard kernel depends on regularization schemes
unavoidably, which can, nevertheless, be regarded as part of the
factorization-scheme dependence. This dependence, usually
minimized by adhering to a fixed prescription for deriving hard
kernels, does not cause a problem. The removal of the light-cone
singularities from wave functions and the gauge invariance of hard
kernels are the two essential ingredients for making physical
predictions from $k_T$ factorization theorem.

We shall demonstrate that the higher-order quark diagrams for
$\pi\gamma^*\to\gamma$ generate two types of double logarithms,
$\ln^2(Q^2/k_T^2)$ and $\ln^2 x$, $Q^2$ being the large momentum
transfer squared, from the loop correction to the virtual photon
vertex. The former does not appear in collinear factorization
theorem, but the latter does \cite{NLO,NNLO}. It is found that the
effective diagrams reproduce the same double logarithm
$\ln^2(Q^2/k_T^2)$, which is then absorbed into the pion wave
function, and organized by $k_T$ resummation \cite{BS,LY1,CS,MR}.
The remaining double logarithm $\ln^2x$ can be absorbed into the
jet function, and organized by the threshold resummation
\cite{UL}. Eventually, the hard kernel is free of any double
logarithm, and its perturbative expansion is improved. It will be
shown that the NLO corrections are only few-percent for the pion
transition form factor involved in the scattering process
$\pi\gamma^*\to \gamma$, if setting the factorization scale to the
momentum transfer.

In Sec.~II we calculate the $O(\alpha_s)$ quark diagrams from full
QCD, the $O(\alpha_s)$ effective diagrams for the pion wave
function, and the $O(\alpha_s)$ jet function, and then take their
difference to obtain the $O(\alpha_s)$ hard kernel for
$\pi\gamma^*\to\gamma$ in $k_T$ factorization theorem. The gauge
invariance of the $k_T$-dependent hard kernel is proven to all
orders in $\alpha_s$ by induction in Sec.~III. Section IV is the
conclusion.

\section{$O(\alpha_s)$ $k_T$ FACTORIZATION}

In this section we set up the framework for computing the hard
kernel for the pion transition form factor in $k_T$ factorization
theorem. The momentum $P_1$ of the pion and the momentum $P_2$ of
the out-going on-shell photon are chosen as
\begin{eqnarray}
P_1 =(P_1^+,0,{\bf 0}_T)\;, \;\;\; P_2 = (0,P_2^-,{\bf
0}_T)\;.\label{ppb}
\end{eqnarray}
The LO quark diagram, in which the anti-quark $\bar q$ carries the
on-shell fractional momentum $k=(xP_1^+,0,{\bf 0}_T)$ and the
internal quark carries $P_2-k$, leads to the amplitude
\begin{eqnarray}
G^{(0)}(x,Q^2)=\frac{tr[\not \epsilon (\not P_2-\not k) \gamma_\mu
\not P_1\gamma_5]}{(P_2-k)^2}=-\frac{tr[\not \epsilon \not P_2
\gamma_\mu \not P_1\gamma_5]}{x Q^2}\;, \label{p1a}
\end{eqnarray}
with the leading spin structure $\not P_1\gamma_5$ of the pion and
$Q^2\equiv 2P_1\cdot P_2$. We have suppressed other constant
factors, such as the electric charge, the color number, and the
pion decay constant, which are irrelevant in the following
discussion.

The trivial factorization of Eq.~(\ref{p1a}) reads \cite{NL2},
\begin{eqnarray}
G^{(0)}(x,Q^2)&=&\int dx'd^2k'_T\Phi^{(0)}(x;x',k'_T)
H^{(0)}(x',Q^2,k'_T)\;,\nonumber\\
\Phi^{(0)}(x;x',k'_T)&=& \delta(x-x')\delta({\bf k}'_T)\;,
\nonumber\\
H^{(0)}(x,Q^2,k_T)&=&- \frac{tr[\not \epsilon \not P_2 \gamma_\mu
\not P_1\gamma^5]}{x Q^2+k_T^2}\;. \label{h0p}
\end{eqnarray}
Once we concentrate on the small $x$ region, the treatment of the
parton $k_T$ differs from that in collinear factorization theorem:
$k_T^2$ in the denominator of Eq.~(\ref{h0p}) is not small
compared to $xQ^2$, and the internal quark propagator should not
be expanded into a power series in $k_T^2$ \cite{TLS,CKL}. $k_T$
in the numerator, being power-suppressed by $1/Q$, is combined
with three-parton meson wave functions to form a gauge-invariant
set of higher-twist contributions as in collinear factorization
theorem. This special treatment of the parton $k_T$ characterizes
the distinction between $k_T$ and collinear factorizations
\cite{LRev}. Because of the zeroth-order wave function
$\Phi^{(0)}\propto\delta({\bf k}'_T)$, the LO hard kernel
$H^{(0)}$ does not depend on the parton transverse momentum
actually.

The $O(\alpha_s)$ quark diagrams corresponding to Eq.~(\ref{p1a})
from full QCD are displayed in Fig.~\ref{fig1}, in which the upper
line represents the $q$ quark. The factorization of the collinear
divergences from these radiative corrections is referred to
\cite{NL2}:
\begin{eqnarray}
G^{(1)}(x,Q^2)&=&\int dx' d^2k'_T
\left[\Phi^{(1)}(x;x',k'_T)H^{(0)}(x',Q^2,k'_T)
+\Phi^{(0)}(x;x',k'_T)H^{(1)}(x',Q^2,k'_T)\right]\;,
\label{pg11}
\end{eqnarray}
where the $O(\alpha_s)$ effective diagrams $\Phi^{(1)}$ are
defined by the leading-twist quark-level wave function
\cite{NL2,L1}
\begin{eqnarray}
\Phi(x;x',k'_T)=\int\frac{dy^-}{2\pi
i}\frac{d^2y_T}{(2\pi)^2}e^{-ix'P_1^+ y^-+i{\bf k}'_T\cdot {\bf
y}_T}\langle 0|{\bar q}(y) W_y(n)^{\dag}I_{n;y,0}W_0(n) \not
n_-\gamma_5 q(0)|q(P_1-k)\bar q(k)\rangle\;,\label{de1}
\end{eqnarray}
with $y=(0,y^-,{\bf y}_T)$ being the coordinate of the anti-quark
field $\bar q$, $n_-=(0,1,{\bf 0}_T)$ a null vector along $P_2$,
and $|q(P_1- k)\bar q(k)\rangle$ the leading Fock state of the
pion.

The factor $W_y(n)$ with $n^2\not=0$ denotes the Wilson line
operator,
\begin{eqnarray}
\label{eq:WL.def} W_y(n) = P \exp\left[-ig \int_0^\infty d\lambda
n\cdot A(y+\lambda n)\right]\;.
\end{eqnarray}
The two Wilson lines $W_y(n)$ and $W_0(n)$ are connected by a link
$I_{n;y,0}$ at infinity in this case \cite{NL2,BJY}.
Equation~(\ref{de1}) contains additional collinear divergences
from the region with a loop momentum parallel to $n_-$, as the
Wilson line direction approaches the light cone, ie., as $n\to
n_-$ \cite{Co03}. It will be shown that $n^2$ serves as an
infrared regulator for the light-cone singularities, and that the
wave function depends on the additional scale $\zeta^2\equiv 4(n
\cdot P_1)^2/|n^2|$, ie., on the external kinematic variable.
Besides, $\Phi$ also depends on the factorization scale $\mu_{\rm
f}$, which is not shown explicitly. Note that Eq.~(\ref{de1}) does
not reduce to the distribution amplitude in collinear
factorization theorem directly, when integrated over $k_T$, but a
convolution of a hard kernel with the distribution amplitude
\cite{Li98}.

With one-gluon exchange, the outgoing partons from $\Phi^{(1)}$,
ie., the partons participating the hard scattering, carry the
transverse momenta, so that $H^{(0)}$ in Eq.~(\ref{pg11}) depends
on $k'_T$ nontrivially in the first-order factorization. Being
convoluted with $\Phi^{(0)}$, the partons entering the NLO hard
kernel $H^{(1)}$ are still on-shell. To acquire the nontrivial
$k_T$ dependence, $H^{(1)}$ must be convoluted with the
higher-order wave functions $\Phi^{(i)}$, $i\ge 1$: the gluon
exchanges in $\Phi^{(i)}$ render the incoming partons of
$H^{(1)}$, ie., the incoming partons of the quark diagrams
$G^{(1)}$ and the effective diagrams $\Phi^{(1)}$ off-shell by
$k_T^2$ \cite{NL2}. We thus derive $H^{(1)}(x,Q^2,k_T)$ according
to the formula
\begin{eqnarray}
H^{(1)}(x,Q^2,k_T)&=&G^{(1)}(x,Q^2,k_T)-\int dx' d^2k'_T
\Phi^{(1)}(x,k_T;x',k'_T)H^{(0)}(x',Q^2,k'_T)\;, \label{pa1}
\end{eqnarray}
where $\Phi^{(1)}(x,k_T;x',k'_T)$ is defined by Eq.~(\ref{de1})
but with the $\bar q$ quark momentum $k=(xP_1^+,0,{\bf k}_T)$. As
stated in the Introduction, the gauge dependences of $G^{(1)}$ and
$\Phi^{(1)}$ cancel in the above expression, such that
$H^{(1)}(x,Q^2,k_T)$ turns out to be gauge-invariant.

\subsection{Quark Diagrams}

\begin{figure}[tb]
\begin{center}
\includegraphics[scale=0.9]{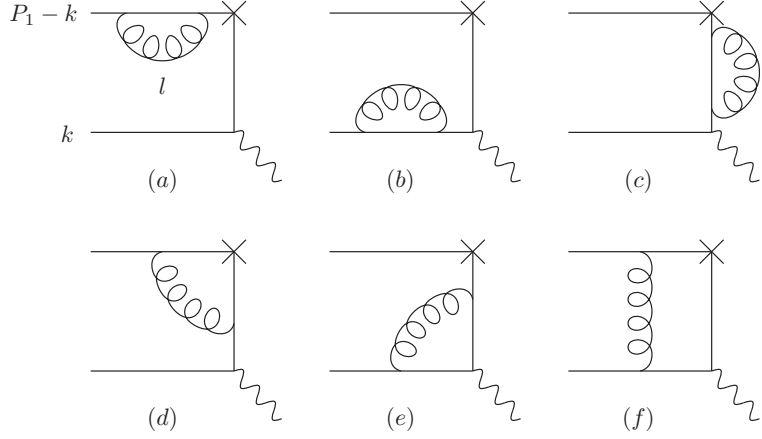}
\caption{$O(\alpha_s)$ quark diagrams for $\pi\gamma^*\to\gamma$
with $\times$ representing the virtual photon vertex.
}\label{fig1}
\end{center}
\end{figure}

The loop integrals associated with the $O(\alpha_s)$ quark
diagrams in Figs.~\ref{fig1}(a)-(f), where the $\bar q$ quark
carries the momentum $k=(xP_1^+,0,{\bf k}_T)$ and the $q$ quark
carries $\bar k\equiv P_1-k$, are written, in the Feynman gauge,
as
\begin{eqnarray}
G^{(1)}_a(x,Q^2,k_T)&=& \frac{-i}{2}g^2C_F \mu^{2\epsilon}
\int\frac{d^{4-2\epsilon}l}{(2\pi)^{4-2\epsilon}}tr\left[
\not\epsilon\frac{\not P_2-\not k}{(P_2-k)^2}\gamma_\mu\frac{\not
\bar k}{\bar k^2}\gamma^\nu \frac{\not \bar k-\not l}{(\bar
k-l)^2}\gamma_\nu\not P_1\gamma_5\right] \frac{1}{l^2}\;,
\label{pga}\\
G^{(1)}_b(x,Q^2,k_T)&=& \frac{-i}{2}g^2C_F \mu^{2\epsilon}
\int\frac{d^{4-2\epsilon}l}{(2\pi)^{4-2\epsilon}}
tr\left[\gamma^\nu \frac{\not k-\not l}{(k-l)^2}\gamma_\nu
\frac{\not k}{k^2}\not\epsilon\frac{\not P_2-\not
k}{(P_2-k)^2}\gamma_\mu\not P_1 \gamma_5\right]
\frac{1}{l^2}\;, \label{pgb}\\
G^{(1)}_c(x,Q^2,k_T)&=&-ig^2C_F \mu^{2\epsilon}
\int\frac{d^{4-2\epsilon}l}{(2\pi)^{4-2\epsilon}}tr\left[
\not\epsilon\frac{\not P_2-\not k}{(P_2-k)^2}\gamma^\nu\frac{\not
P_2-\not k-\not l}{(P_2-k-l)^2}\gamma_\nu\frac{\not P_2-\not
k}{(P_2-k)^2}\gamma_\mu \not P_1 \gamma_5\right]
\frac{1}{l^2}\;,\label{pgc}\\
G^{(1)}_d(x,Q^2,k_T)&=&-ig^2C_F \mu^{2\epsilon}
\int\frac{d^{4-2\epsilon}l}{(2\pi)^{4-2\epsilon}}tr\left[
\not\epsilon\frac{\not P_2-\not k}{(P_2-k)^2}\gamma^\nu \frac{\not
P_2-\not k+\not l}{(P_2-k+l)^2} \gamma_\mu \frac{\not \bar k+\not
l}{(\bar k+l)^2}\gamma_\nu \not P_1\gamma_5\right]
\frac{1}{l^2}\;,\label{pgd}\\
G^{(1)}_e(x,Q^2,k_T)&=&ig^2C_F \mu^{2\epsilon}
\int\frac{d^{4-2\epsilon}l}{(2\pi)^{4-2\epsilon}}
tr\left[\gamma_\nu \frac{\not k-\not l}{(k-l)^2}
\not\epsilon\frac{\not P_2-\not k+\not
l}{(P_2-k+l)^2}\gamma^\nu\frac{\not P_2-\not
k}{(P_2-k)^2}\gamma_\mu\not P_1\gamma_5\right]
\frac{1}{l^2}\;,\label{pge}\\
G^{(1)}_f(x,Q^2,k_T)&=&ig^2C_F \mu^{2\epsilon}
\int\frac{d^{4-2\epsilon}l}{(2\pi)^{4-2\epsilon}}tr\left[
\gamma^\nu \frac{\not k-\not l}{(k-l)^2}\not\epsilon\frac{\not
P_2-\not k+\not l}{(P_2-k+l)^2} \gamma_\mu \frac{\not \bar k+\not
l}{(\bar k+l)^2}\gamma_\nu \not P_1\gamma_5\right]
\frac{1}{l^2}\;.\label{pgf}
\end{eqnarray}
The coefficients $1/2$ in Eqs.~(\ref{pga}) and (\ref{pgb}) arise
from the definition of the self-energy corrections to external
particles. $C_F$ is a color factor, and $\mu$ the renormalization
scale.

We work in the dimensional reduction \cite{WS79} to simplify the
calculation and to avoid the ambiguity from handling $\gamma_5$ in
arbitrary dimensions. The results for the self-energy corrections
are
\begin{eqnarray}
G^{(1)}_a(x,Q^2,k_T)&=&-\frac{\alpha_s}{8\pi}C_F\left(\frac{1}{\epsilon}
+\ln\frac{4\pi\mu^2}{k_T^2e^{\gamma_E}}+2\right)
H^{(0)}(x,Q^2,k_T)\;,\label{pga1}\\
G^{(1)}_b(x,Q^2,k_T)&=&
-\frac{\alpha_s}{8\pi}C_F\left(\frac{1}{\epsilon}
+\ln\frac{4\pi\mu^2}{k_T^2 e^{\gamma_E}}+2\right)H^{(0)}(x,Q^2,k_T)\;, \label{pgb1}\\
G^{(1)}_c(x,Q^2,k_T)&=&-\frac{\alpha_s}{4\pi}C_F\left(\frac{1}{\epsilon}
+\ln\frac{4\pi\mu^2
e^{-\gamma_E}}{xQ^2+k_T^2}+2\right)H^{(0)}(x,Q^2,k_T)\;,\label{pgc1}
\end{eqnarray}
where $1/\epsilon$ denotes the ultraviolet pole, and $\gamma_E$ is
the Euler constant. Since the external partons are off-shell by
$k_T^2$, the collinear divergences in Figs.~\ref{fig1}(a) and 1(b)
are represented by the infrared logarithms $\ln k_T^2$ in
Eqs.~(\ref{pga1}) and (\ref{pgb1}), respectively. The internal
quark in Fig.~\ref{fig1}(c) is off-shell by the invariant mass
squared $xQ^2+k_T^2$, which then replaces the argument $k_T^2$ in
the infrared logarithm.

In the small $x$ region we drop terms suppressed by powers of $x$
or $k_T^2/Q^2$. The loop correction to the virtual photon vertex
gives
\begin{eqnarray}
G^{(1)}_d(x,Q^2,k_T)&=&\frac{\alpha_s}{4\pi}C_F\Bigg(\frac{1}{\epsilon}
+\ln\frac{4\pi\mu^2}{k_T^2e^{\gamma_E}}
-2\ln\frac{Q^2}{k_T^2}\ln\frac{Q^2}{xQ^2+k_T^2}
\nonumber\\
&&+2\ln\frac{Q^2}{xQ^2+k_T^2}+\ln\frac{Q^2}{k_T^2}-\frac{2\pi^2}{3}+\frac{3}{2}
\Bigg)H^{(0)}(x,Q^2,k_T)\;.\label{pgd1}
\end{eqnarray}
At small $x$ the $q$ quark in Fig.~\ref{fig1}(d) is energetic,
implying the existence of the collinear logarithmic enhancement
$\ln(Q^2/k_T^2)$, and the internal quark is close to mass shell,
implying the soft enhancement $\ln[Q^2/(xQ^2+k_T^2)]$. Their
overlap then leads to the double logarithm
$\ln(Q^2/k_T^2)\ln[Q^2/(xQ^2+k_T^2)]$ in Eq.~(\ref{pgd1}). In the
region with $x\sim O(1)$, the internal quark becomes off-shell by
$O(Q^2)$, the soft enhancement disappears as
$\ln[Q^2/(xQ^2+k_T^2)]\sim O(1)$, and the double logarithm reduces
to a single logarithm. The result of $G^{(1)}_d$ clearly exhibits
the transition of the double logarithm in the small $x$ region to
the single logarithm in the large $x$ region.

The above double logarithm deserves more discussion, which can be
reexpressed as
\begin{eqnarray}
-2\ln\frac{Q^2}{k_T^2}\ln\frac{Q^2}{xQ^2+k_T^2}=-\ln^2\frac{Q^2}{k_T^2}
-\ln^2\frac{Q^2}{xQ^2+k_T^2}+\ln^2\frac{xQ^2+k_T^2}{k_T^2}\;.\label{double}
\end{eqnarray}
The first term is known as the Sudakov logarithm \cite{BS,CS},
which will be absorbed into the pion wave function as stated
before. The Sudakov effect from resumming this double logarithm
suppresses the contribution from the small $k_T$ region, ie., the
region with a large impact parameter \cite{LS}. The second term
exists even in collinear factorization theorem without taking into
account $k_T$ \cite{NLO,ASY}, $\ln[Q^2/(xQ^2+k_T^2)]\sim \ln^2x$,
which can not be factorized into the pion wave function. This
threshold logarithm is important at small $x$, where the internal
quark approaches mass shell. Hence, a jet function has been
introduced to absorb $\ln^2x$, and its resummation effect
suppresses contributions from the small $x$ region \cite{UL}. The
third term, being of $O(1)$, does not require an all-order
organization.

The loop correction to the out-going on-shell photon vertex is
written as
\begin{eqnarray}
G^{(1)}_e(x,Q^2,k_T)&=&\frac{\alpha_s}{4\pi}C_F\left(\frac{1}{\epsilon}
+\ln\frac{4\pi\mu^2}{k_T^2e^{\gamma_E}}
+\ln\frac{xQ^2+k_T^2}{k_T^2}+\frac{3}{2}\right)
H^{(0)}(x,Q^2,k_T)\;,
\end{eqnarray}
which does not contain a double logarithm for the following
reason. In the large $x$ region the internal quark is off-shell by
$O(Q^2)$, and the soft enhancement disappears. In the small $x$
region the $\bar q$ quark becomes soft, and the associated
collinear enhancement is diminished by the limited phase space for
the loop momentum. Therefore, there is a lack of overlap of the
collinear and soft enhancements, and only the $O(1)$ single
logarithm exists.

At last, the evaluation of the box diagram Fig.~\ref{fig1}(f) is
simple, giving a power-suppressed contribution at small $x$. In
the region with $x\sim O(1)$, ie., $k^+\sim O(Q)$, the internal
quark in Fig.~\ref{fig1}(f) is off-shell by $1/[P_2\cdot
(k-l)]\sim 1/Q^2$ for either a collinear loop momentum $l^+\sim
O(Q)$ or an ultraviolet loop momentum $l^\mu\sim O(Q)$, the same
as $1/(P_2\cdot k)\sim 1/Q^2$ in the LO amplitude. Namely, the
radiative correction from the box diagram does not change the LO
power-law behavior, and its contribution is finite. In the region
with small $x\sim O(\Lambda)$, $\Lambda$ being a hadronic scale,
the LO amplitude scales like $1/(P_2\cdot k)\sim 1/(Q\Lambda)$,
while the internal quark in Fig.~\ref{fig1}(f) remains off-shell
by $1/[P_2\cdot (k-l)]\sim 1/Q^2$ for either collinear or
ultraviolet $l$. Thus the contribution from the box diagram
becomes power-suppressed and negligible, and we have
$G^{(1)}_f(x,Q^2,k_T)=0$ at leading power. The above observation
is consistent with the corresponding NLO analysis in collinear
factorization theorem \cite{NLO}, which indicates the vanishing of
the box-diagram contribution in the small $x$ region explicitly.

The sum of the radiative corrections from the quark diagrams
Figs.~\ref{fig1}(a)-(f) gives
\begin{eqnarray}
G^{(1)}(x,Q^2,k_T)&=&\sum_{i=a}^{f}G^{(1)}_i(x,Q^2,k_T)\nonumber\\
&=&-\frac{\alpha_s}{4\pi}C_F\Bigg(
2\ln\frac{Q^2}{k_T^2}\ln\frac{Q^2}{xQ^2+k_T^2}
-3\ln\frac{Q^2}{k_T^2}+1+\frac{2\pi^2}{3}
\Bigg)H^{(0)}(x,Q^2,k_T)\;.\label{pgt}
\end{eqnarray}
It is observed that all the ultraviolet poles cancel and the $\mu$
dependence disappears completely, a consequence of the
conservation of the current that defines the pion transition form
factor. It will be demonstrated in the next subsection that the
effective diagrams for the pion wave function generate the same
infrared logarithms $\ln k_T^2$.

\subsection{Effective Diagrams}

\begin{figure}[tb]
\begin{center}
\includegraphics[scale=0.8]{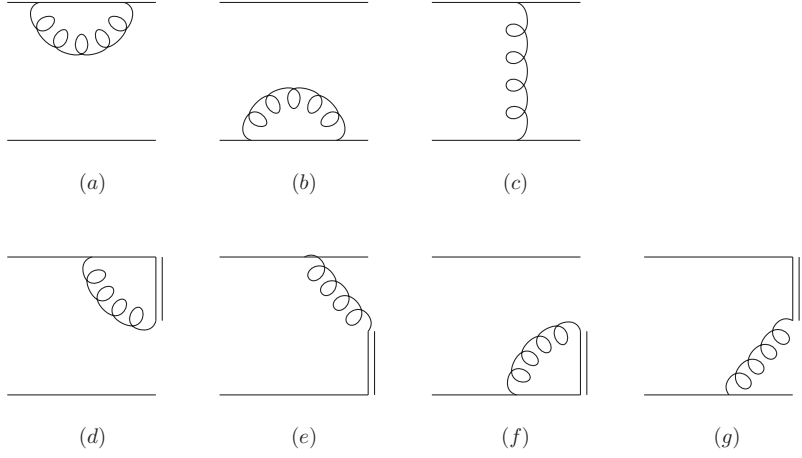}
\caption{$O(\alpha_s)$ effective diagrams for the pion wave
function.}\label{fig2}
\end{center}
\end{figure}

We first explain the appearance of the nonphysical light-cone
divergences in the naive definition for $k_T$-dependent hadron
wave functions. To factor out the collinear gluons in Figs.~1(d)
and 1(e), the following approximation for the product of the two
internal quark propagators has been employed \cite{L1},
\begin{eqnarray}
\frac{2P_2^\nu}{(P_2-k)^2(P_2-k+l)^2}
\approx\frac{n_-^\nu}{n_-\cdot l}\biggl[-\frac{1}{xQ^2+k_T^2}
+\frac{1}{(x-l^+/P_1^+)Q^2+|{\bf k}_T-{\bf l}_T|^2}\biggr]\;,
\label{pi}
\end{eqnarray}
where $2P_2^\nu$ comes from the contraction of $\not P_2$ and
$\gamma^\nu$ in the numerators of Eqs.~(\ref{pgd}) and
(\ref{pge}). The factor $n_-^\nu/n_-\cdot l$ is exactly the
Feynman rule associated with the Wilson line along the light cone,
which is necessary for the gauge invariance of the nonlocal matrix
element in the pion wave function. The first (second) term in the
above splitting corresponds to the case without (with) the loop
momentum $l$ flowing through the hard scattering. It is easy to
see that the right-hand side of Eq.~(\ref{pi}) is well-defined in
the $n_-\cdot l=l^+\to 0$ limit, if the transverse momenta $k_T^2$
and $|{\bf k}_T-{\bf l}_T|^2$ are dropped. That is, collinear
factorization can be made gauge-invariant and free of the
light-cone singularities. However, singularities from $l^+\to 0$
are developed, when the transverse momenta are included, implying
that the factorization of collinear gluons should be performed
more carefully in $k_T$ factorization theorem. This is the reason
the naive definition is modified into Eq.~(\ref{de1}) with the
non-light-like vector $n$, which makes finite $n\cdot l$ as
$l^+\to 0$.

The explicit expressions for the $O(\alpha_s)$ effective diagrams
displayed in Fig.~\ref{fig2}(a)-(g) are written, following
Eq.~(\ref{de1}), as
\begin{eqnarray}
\Phi^{(1)}_{a}(x,k_T;x',k'_T) &=& -\frac{i}{8}g^2C_F \mu_{\rm
f}^{2\epsilon}
\int\frac{d^{4-2\epsilon}l}{(2\pi)^{4-2\epsilon}}tr\left[
\gamma_5\not n_-\frac{\not \bar k}{\bar k^2}\gamma^\nu \frac{\not
\bar k-\not l}{(\bar k-l)^2}\gamma_\nu \not n_+\gamma_5\right]
\frac{1}{l^2}
\nonumber\\
& &\times \delta(x-x')\delta({\bf k}_T-{\bf k}'_T)\;,
\label{pka}\\
\Phi^{(1)}_b(x,k_T;x',k'_T)&=& -\frac{i}{8}g^2C_F \mu_{\rm
f}^{2\epsilon}
\int\frac{d^{4-2\epsilon}l}{(2\pi)^{4-2\epsilon}}tr\left[
\gamma^\nu \frac{\not k-\not l}{(k-l)^2}\gamma_\nu \frac{\not
k}{k^2}\gamma_5\not n_- \not n_+\gamma_5\right]\frac{1}{l^2}
\nonumber\\
& &\times \delta(x-x')\delta({\bf k}_T-{\bf k}'_T)\;,
\label{pkb}\\
\Phi^{(1)}_c(x,k_T;x',k'_T)&=&\frac{i}{4}g^2C_F \mu_{\rm
f}^{2\epsilon}
\int\frac{d^{4-2\epsilon}l}{(2\pi)^{4-2\epsilon}}\left[ \gamma^\nu
\frac{\not k-\not l}{(k-l)^2}\gamma_5\not n_- \frac{\not \bar
k+\not l}{(\bar k+l)^2}\gamma_\nu\not
n_+\gamma_5\right]\frac{1}{l^2}
\nonumber\\
& &\times \delta\left(x-x'-\frac{l^+}{P_1^+}\right) \delta({\bf
k}_T-{\bf k}'_T-{\bf l}_T)\;, \label{pkc}\\
\Phi^{(1)}_{d}(x,k_T;x',k'_T)&=&-\frac{i}{4}g^2C_F\mu_{\rm
f}^{2\epsilon} \int\frac{d^{4-2\epsilon}l}{(2\pi)^{4-2\epsilon}}
tr\left[ \gamma_5\not n_-\frac{\not \bar k+\not l}{(\bar
k+l)^2}\gamma_\nu\not n_+\gamma_5\right]
\frac{1}{l^2}\frac{n^\nu}{n\cdot l}\nonumber\\
& &\times \delta(x-x')\delta({\bf k}_T-{\bf k}'_T)\;, \label{pkd}\\
\Phi^{(1)}_{e}(x,k_T;x',k'_T)&=&\frac{i}{4}g^2C_F\mu_{\rm
f}^{2\epsilon} \int\frac{d^{4-2\epsilon}l}{(2\pi)^{4-2\epsilon}}
tr\left[ \gamma_5\not n_-\frac{\not \bar k+\not l}{(\bar
k+l)^2}\gamma_\nu\not n_+\gamma_5\right]
\frac{1}{l^2}\frac{n^\nu}{n\cdot l}\nonumber\\
& &\times \delta\left(x-x'-\frac{l^+}{P_1^+}\right) \delta({\bf
k}_T-{\bf k}'_T-{\bf l}_T)\;, \label{pke}\\
\Phi^{(1)}_f(x,k_T;x',k'_T)&=& \frac{i}{4}g^2C_F\mu_{\rm
f}^{2\epsilon} \int\frac{d^{4-2\epsilon}l}{(2\pi)^{4-2\epsilon}}
tr\left[\gamma_\nu\frac{\not k-\not l}{(k-l)^2} \gamma_5\not n_-
\not n_+\gamma_5\right]\frac{1}{l^2}\frac{n^\nu}{n\cdot l}
\nonumber\\
& &\times\delta(x-x')\delta({\bf k}_T-{\bf k}'_T)\;, \label{pkf}\\
\Phi^{(1)}_g(x,k_T;x',k'_T)&=& -\frac{i}{4}g^2C_F\mu_{\rm
f}^{2\epsilon} \int\frac{d^{4-2\epsilon}l}{(2\pi)^{4-2\epsilon}}
tr\left[\gamma_\nu\frac{\not k-\not l}{(k-l)^2} \gamma_5\not n_-
\not n_+\gamma_5\right]\frac{1}{l^2}\frac{n^\nu}{n\cdot l}
\nonumber\\
& &\times \delta\left(x-x'-\frac{l^+}{P_1^+}\right)\delta({\bf
k}_T-{\bf k}'_T-{\bf l}_T)\;, \label{pkg}
\end{eqnarray}
where $n_+=(1,0,{\bf 0}_T)$ is a null vector along the pion
momentum $P_1$, and the arguments $\mu_{\rm f}$ and $\zeta^2$ of
$\Phi^{(1)}$ are not exhibited for brevity. Note that the
zeroth-order wave function is given by
$\Phi^{(0)}=\delta(x-x')\delta({\bf k}_T-{\bf k}'_T)$ here.

We compute the convolution of $\Phi^{(1)}$ with the LO hard kernel
$H^{(0)}$ in Eq.~(\ref{h0p}) over the integration variables $x'$
and $k'_T$, denoted by $\otimes$ below:
\begin{eqnarray}
\Phi^{(1)}_{i}\otimes H^{(0)}\equiv \int dx' d^2k'_T
\Phi^{(1)}_i(x,k_T;x',k'_T)H^{(0)}(x',Q^2,k'_T)\;.
\end{eqnarray}
The self-energy corrections in Figs.~\ref{fig2}(a) and
\ref{fig2}(b) are similar to the quark diagrams in
Figs.~\ref{fig1}(a) and \ref{fig1}(b), respectively, and the
results are
\begin{eqnarray}
\Phi^{(1)}_{a}\otimes
H^{(0)}&=&-\frac{\alpha_s}{8\pi}C_F\left(\frac{1}{\epsilon}
+\ln\frac{4\pi\mu_{\rm
f}^2}{k_T^2e^{\gamma_E}}+2\right)H^{(0)}(x,Q^2,k_T)\;,
\label{pwa}\\
\Phi^{(1)}_{b}\otimes
H^{(0)}&=&-\frac{\alpha_s}{8\pi}C_F\left(\frac{1}{\epsilon}
+\ln\frac{4\pi\mu_{\rm
f}^2}{k_T^2e^{\gamma_E}}+2\right)H^{(0)}(x,Q^2,k_T)\;.\label{pwb}
\end{eqnarray}
Similarly, the contribution from the box diagram
Fig.~\ref{fig2}(c) is power-suppressed in the small $x$ region,
and we have $\Phi^{(1)}_{c}\otimes H^{(0)}=0$.

When evaluating Eqs.~(\ref{pkd})-(\ref{pkg}), the sign of the plus
component $n^+$ of the vector $n$ is arbitrary, which could be
positive or negative ($n^-$ has a positive sign, the same as of
$P_2^-$). Choosing $n^+<0$, ie., $n^2<0$ as in
\cite{LS,LY1,KLS,LUY}, Fig.~\ref{fig2}(d) leads, in the small $x$
region, to
\begin{eqnarray}
\Phi^{(1)}_{d}\otimes
H^{(0)}=\frac{\alpha_s}{4\pi}C_F\left(\frac{1}{\epsilon}
+\ln\frac{4\pi\mu_{\rm f}^2}{k_T^2e^{\gamma_E}}
-\ln^2\frac{\zeta^2}{k_T^2}+\ln\frac{\zeta^2}{k_T^2}+2-\frac{\pi^2}{3}
\right)H^{(0)}(x,Q^2,k_T)\;,\label{pwd}
\end{eqnarray}
which reproduces the Sudakov logarithm $\ln^2 (Q^2/k_T^2)$ from
Fig.~\ref{fig1}(d) in Eq.~(\ref{double}), noticing the scale
$\zeta^2=|n^-/n^+|Q^2$. The light-cone divergences are regularized
in the price that the universality of a wave function is lost, for
it depends on the external kinematic variable through $\zeta^2$.
This problem can be alleviated by extracting the evolution in
$\zeta^2$ from Eq.~(\ref{de1}) \cite{Co03}, ie., by resumming
$\ln^2(\zeta^2/k_T^2)$ in Eq.~(\ref{pwd}) into the Sudakov factor
\cite{CS,Li96}. The initial condition of the evolution is
universal, like a distribution amplitude in collinear
factorization theorem. We stress that the Sudakov resummation,
accurate up to fixed loops, does not remove the $\zeta^2$
dependence of a wave function completely. That is,
nonfactorizability may occur at subleading level in $k_T$
factorization of the pion transition form factor.

The hard kernel associated with $\Phi_e^{(1)}$, ie., the second
term in Eq.~(\ref{pi}), demands the physical range of $l^+$ to be
$-\bar k^+\le l^+\le k^+$, which corresponds to the range of the
parton momentum fraction $1\ge x'\ge 0$. As computing the
convolution of $\Phi_e^{(1)}$ with $H^{(0)}$, this fact should be
taken into account. Moreover, we assume $\zeta^2\sim Q^2$ by
choosing $|n^+|\sim n^-$ to avoid creating the additional large
logarithm $\ln(\zeta^2/Q^2)$. The leading-power expression for
Fig.~\ref{fig2}(e) is then given, in the small $x$ region, by
\begin{eqnarray}
\Phi^{(1)}_{e}\otimes H^{(0)}&=&
\frac{\alpha_s}{4\pi}C_F \ln^2\frac{\zeta^2(xQ^2+k_T^2)}{Q^2k_T^2}
H^{(0)}(x,Q^2,k_T)\;, \label{pwe}
\end{eqnarray}
where terms vanishing with $k_T^2\to 0$ have been dropped. It is
found that Fig.~\ref{fig2}(e) does not generate a large double
logarithm with $\zeta^2\sim Q^2$.

It is interesting to obtain the results corresponding to $n^+>0$
for Figs.~\ref{fig2}(d) and \ref{fig2}(e). One simply analytically
continues Eqs.~(\ref{pwd}) and (\ref{pwe}) into the region with
$n^2>0$ by means of the principle-value prescription,
\begin{eqnarray}
P\left[\ln^2\frac{(n\cdot P_1)^2}{n^2}\right] =\ln^2\frac{(n\cdot
P_1)^2}{|n^2|}-\pi^2\;,\;\;\;\; P\left[\ln\frac{(n\cdot
P_1)^2}{n^2}\right] =\ln\frac{(n\cdot P_1)^2}{|n^2|}\;.
\end{eqnarray}
We then derive from Eq.~(\ref{pwd})
\begin{eqnarray}
\Phi^{(1)}_{d}\otimes
H^{(0)}=\frac{\alpha_s}{4\pi}C_F\left(\frac{1}{\epsilon}
+\ln\frac{4\pi\mu_{\rm f}^2}{k_T^2e^{\gamma_E}}
-\ln^2\frac{\zeta^2}{k_T^2}+\ln\frac{\zeta^2}{k_T^2}+2-\frac{4\pi^2}{3}
\right)H^{(0)}(x,Q^2,k_T)\;,\label{pwd1}
\end{eqnarray}
which can be confirmed by calculating the loop integral in
Eq.~(\ref{pkd}) directly for $n^2>0$. It shows that the choices
$n^2>0$ and $n^2<0$ lead to expressions different only by a
constant term. Because the $n$-dependent double logarithms cancel
in the summation
\begin{eqnarray}
(\Phi^{(1)}_{d}+\Phi^{(1)}_{e})\otimes
H^{(0)}&=&\frac{\alpha_s}{4\pi}C_F\left[\frac{1}{\epsilon}
+\ln\frac{4\pi\mu_{\rm f}^2}{k_T^2e^{\gamma_E}}
-2\ln\frac{\zeta^2}{k_T^2}\ln\frac{Q^2}{xQ^2+k_T^2}
+\ln^2\frac{Q^2}{xQ^2+k_T^2}\right.\nonumber\\
& &\left. +\ln\frac{\zeta^2}{k_T^2} +2-\frac{\pi^2}{3}
\right]H^{(0)}(x,Q^2,k_T)\;,\label{pwde}
\end{eqnarray}
$(\Phi^{(1)}_{d}+\Phi^{(1)}_{e})\otimes H^{(0)}$ does not depend
on the sign of $n^2$ actually.

Applying the variable change $l\to -l$, and the transformation
$n\to -n$ and $k\to \bar k$, Eq.~(\ref{pkf}) becomes identical to
Eq.~(\ref{pkd}). Therefore, the result from Fig.~\ref{fig2}(f) is
the same as of Fig.~\ref{fig2}(d), but with the replacement of
$\bar k\approx P_1$ by $k$, ie., $\zeta$ by $x\zeta$. Keeping
terms which do not vanish with $k_T^2\to 0$, we have
\begin{eqnarray}
\Phi^{(1)}_{f}\otimes
H^{(0)}&=&\frac{\alpha_s}{4\pi}C_F\left(\frac{1}{\epsilon}
+\ln\frac{4\pi\mu_{\rm f}^2}{k_T^2e^{\gamma_E}}
-\ln^2\frac{x^2\zeta^2}{k_T^2}+\ln\frac{x^2\zeta^2}{k_T^2}+2-\frac{\pi^2}{3}
\right)H^{(0)}(x,Q^2,k_T)\;,\label{pwf}
\end{eqnarray}
where the double logarithm, being large in the region of $x\sim
O(1)$, attenuates with the decrease of $x$. It should disappear,
after combined with the contribution from Fig.~\ref{fig2}(g),
since such a double logarithm is absent in the corresponding quark
diagram Fig.~\ref{fig1}(e) in any region of $x$. The same variable
transformation relating $\Phi^{(1)}_{f}$ to $\Phi^{(1)}_{d}$ is
not applicable to $\Phi^{(1)}_{g}$, for the latter involves the
nontrivial convolution with $H^{(0)}$. Hence,
$\Phi^{(1)}_{g}\otimes H^{(0)}$ is expected to have an expression
different from $\Phi^{(1)}_{e}\otimes H^{(0)}$. Retaining terms
which are finite as $k_T\to 0$, Fig.~\ref{fig2}(g) leads, in the
small $x$ region with $xQ^2\gg x^2\zeta^2$, to
\begin{eqnarray}
\Phi^{(1)}_{g}\otimes H^{(0)}&=&\frac{\alpha_s}{4\pi}C_F
\ln^2\frac{x^2\zeta^2}{k_T^2}H^{(0)}(x,Q^2,k_T)\;.\label{pwg}
\end{eqnarray}
The cancellation of the double logarithms in the summation of
Eqs.~(\ref{pwf}) and (\ref{pwg}) is obvious. For a similar reason,
$(\Phi^{(1)}_{f}+\Phi^{(1)}_{g})\otimes H^{(0)}$ is independent of
the sign of $n^2$.

Summing all the above $O(\alpha_s)$ quark-level wave functions, we
derive
\begin{eqnarray}
\Phi^{(1)}\otimes H^{(0)}&=&\sum_{i=a}^{g}\Phi^{(1)}_i\otimes
H^{(0)}\nonumber\\
&=&\frac{\alpha_s}{4\pi}C_F\left(\frac{1}{\epsilon}
+\ln\frac{4\pi\mu_{\rm f}^2}{k_T^2e^{\gamma_E}}
-\ln^2\frac{\zeta^2}{k_T^2}+\ln^2\frac{\zeta^2(xQ^2+k_T^2)}{Q^2k_T^2}
\right.\nonumber\\
& &\left. +\ln\frac{\zeta^2}{k_T^2}+\ln\frac{x^2\zeta^2}{k_T^2}
+2-\frac{2\pi^2}{3} \right)H^{(0)}(x,Q^2,k_T)\;.\label{ppt}
\end{eqnarray}
In contrast to Eq.~(\ref{pgt}), which is independent of the
renormalization scale $\mu$, the above expression depends on the
factorizations scale $\mu_{\rm f}$. The Sudakov resummation and
the renormalization-group method can be applied to organize the
logarithms $\ln^2(\zeta^2/k_T^2)$ and $\ln(\mu^2_{\rm f}/k_T^2)$
to all orders, respectively \cite{LY1}.

\subsection{$O(\alpha_s)$ Hard Kernel}

\begin{figure}[tb]
\begin{center}
\includegraphics[scale=0.9]{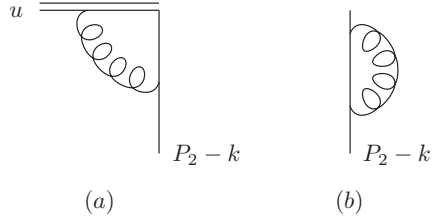}
\caption{$O(\alpha_s)$ diagrams for the jet function.}\label{fig3}
\end{center}
\end{figure}

We renormalize Eq.~(\ref{ppt}) in the modified minimal subtraction
scheme, and then take the difference of Eqs.~(\ref{pgt}) and
(\ref{ppt}) to obtain the $O(\alpha_s)$ hard kernel for the pion
transition form factor. It is easy to find that the hard kernels
$H^{(1)}_{a,b}\equiv G^{(1)}_{a,b}-\Phi^{(1)}_{a,b}\otimes
H^{(0)}$, $H^{(1)}_c\equiv G^{(1)}_c$, $H^{(1)}_d\equiv
G^{(1)}_d-(\Phi^{(1)}_{d}+\Phi^{(1)}_{e})\otimes H^{(0)}$,
$H^{(1)}_e\equiv G^{(1)}_e-(\Phi^{(1)}_{f}+\Phi^{(1)}_{g})\otimes
H^{(0)}$, and $H^{(1)}_f\equiv G^{(1)}_f-\Phi^{(1)}_{c}\otimes
H^{(0)}=0$ associated with Figs.~\ref{fig1}(a)-(f) are all free of
the infrared logarithms $\ln k_T^2$ as claimed before. Compared to
\cite{MW0607}, we do not need the additional soft function $S$ to
achieve this cancellation. The difference is that the self-energy
corrections to the Wilson lines have been included into the set of
effective diagrams for the pion wave function in \cite{MW0607}.
Hence, $S$ must be introduced to remove these artificially
included infrared divergences. We stress that the self-energy
corrections to the Wilson lines do not exist, because such
diagrams are not generated in the derivation of factorization
theorem using the diagrammatic approach \cite{NL2}. This
observation is consistent with the postulation that the gauge
fields appearing in the Wilson lines in Eq.~(\ref{eq:WL.def}) are
regarded as bare fields \cite{Co03}.

After subtracting the effective diagrams from the quark diagrams,
the resultant hard kernel depends on the factorization scheme that
defines the renormalization of Eq.~(\ref{ppt}). The quark diagrams
do not have such a scheme dependence as shown in Eq.~(\ref{pgt}).
When making a physical prediction from factorization theorem, one
convolutes the hard kernel with a model for the pion wave function
(not with the effective diagrams), so that the scheme dependence
in the hard kernel remains. As stated in the Introduction, the
scheme dependence of physical predictions is usually minimized by
adhering to a fixed prescription for deriving hard kernels, which
will be elucidated below. The sum of the $O(\alpha_s)$ hard
kernels is written as
\begin{eqnarray}
H^{(1)}(x,Q^2,k_T)&=&\sum_{i=a}^{f}H^{(1)}_i(x,Q^2,k_T)\nonumber\\
&=&\frac{\alpha_s}{4\pi}C_F\Bigg(-\ln\frac{\mu_{\rm
f}^2}{xQ^2+k_T^2}
+2\ln\frac{\zeta^2}{Q^2}\ln\frac{Q^2}{xQ^2+k_T^2}
-\ln^2\frac{Q^2}{xQ^2+k_T^2} \nonumber\\
& & +2\ln\frac{Q^2}{x\zeta^2}+\ln\frac{Q^2}{xQ^2+k_T^2} -3
\Bigg)H^{(0)}(x,Q^2,k_T)\;.\label{pht}
\end{eqnarray}
The Sudakov logarithm $\ln^2(Q^2/k_T^2)$ in Eq.~(\ref{double}) for
$G_d^{(1)}$ has been cancelled by that in Eq.~(\ref{pwd}) for
$\Phi_d^{(1)}\otimes H^{(0)}$, but the threshold logarithm
$\ln^2[Q^2/(xQ^2+k_T^2)]$ remains in $H^{(1)}$. The large
threshold logarithm can be absorbed into a jet function \cite{UL},
so that the pertubative expansion of the hard kernel is further
improved. At small $x$, a collinear enhancement arises from the
region with the loop momentum parallel to the internal quark
momentum $P_2-k\approx P_2$. To factorize this collinear gluon
into the jet function, we replace the $q$ quark by the eikonal
line in some direction $u$ \cite{Li96,LL} as shown in
Fig.~\ref{fig3}(a). Similarly, we choose $u^2\not=0$ to avoid
other infrared divergences, such as those from $l$ parallel to
$P_1$, which have been absorbed into the pion wave function.
Including the self-energy correction to the internal quark,
Fig.~\ref{fig3}(b), we arrive at the complete set of diagrams for
the jet function at $O(\alpha_s)$.

Figure~\ref{fig3} has been evaluated in \cite{UL}, focusing only
on the double-logarithm piece $\ln^2x$. Here we work out the
single-logarithm and constant pieces too. The explicit expression
of the loop integral $J^{(1)}_{a}$ associated with
Fig.~\ref{fig3}(a) is referred to \cite{UL}. We obtain, for
$u^2<0$,
\begin{eqnarray}
J^{(1)}_{a}H^{(0)}=\frac{\alpha_s}{4\pi}C_F\left(\frac{1}{\epsilon}
+\ln\frac{4\pi\mu^2 e^{-\gamma_E}}{xQ^2+k_T^2}
-\ln^2\frac{\zeta_u^2}{xQ^2+k_T^2}
+\ln\frac{\zeta_u^2}{xQ^2+k_T^2}+2-\frac{\pi^2}{3}
\right)H^{(0)}(x,Q^2,k_T) \;,\label{pja}
\end{eqnarray}
with the scale $\zeta_u^2=4(u\cdot P_2)^2/|u^2|$.
Figure~\ref{fig3}(b) gives a result identical to Eq.~(\ref{pgc1})
for Fig.~\ref{fig1}(c):
\begin{eqnarray}
J^{(1)}_{b}H^{(0)}
&=&-\frac{\alpha_s}{4\pi}C_F\left(\frac{1}{\epsilon}
+\ln\frac{4\pi\mu^2
e^{-\gamma_E}}{xQ^2+k_T^2}+2\right)H^{(0)}(x,Q^2,k_T)\;.\label{pjc}
\end{eqnarray}
Note that the sum $J^{(1)}=J^{(1)}_{a}+J^{(1)}_{b}$ is free of
ultraviolet divergences and $\mu$-independent. That is, the
factorization of the jet function does not modify the
renormalization-group behavior of the hard kernel. As expected,
the jet function is characterized by the invariant mass of the
internal quark.

Define $\zeta^2=\nu Q^2$ and $\zeta_u^2=\nu_u Q^2$ with $\nu$ and
$\nu_u$ being constants of $O(1)$. The hard kernel, after
subtracting the $O(\alpha_s)$ jet function, is given by
\begin{eqnarray}
[H/J]^{(1)}(x,Q^2,k_T)&\equiv&
H^{(1)}(x,Q^2,k_T)-J^{(1)}(x,Q^2,k_T)H^{(0)}(x,Q^2,k_T)\nonumber\\
&=&-\frac{\alpha_s}{4\pi}C_F\Bigg[\ln\frac{\mu_{\rm
f}^2}{xQ^2+k_T^2} -2(\ln\nu+\ln\nu_u)\ln\frac{Q^2}{xQ^2+k_T^2}
+2\ln x\nonumber\\
& &3-\frac{\pi^2}{3}-\ln^2\nu_u+\ln\nu_u+2\ln\nu
\Bigg]H^{(0)}(x,Q^2,k_T)\;,\label{phtt}
\end{eqnarray}
in which the double logarithms have been completely removed.
Different values of $\nu$ and $\nu_u$ correspond to different
factorization schemes. Adopting $\nu=1$, ie., $\zeta^2=Q^2$ as in
\cite{LS}, and $\nu_u=1$, Eq.~(\ref{phtt}) reduces to
\begin{eqnarray}
[H/J]^{(1)}(x,Q^2,k_T)
=-\frac{\alpha_s}{4\pi}C_F\Bigg(\ln\frac{\mu_{\rm
f}^2}{xQ^2+k_T^2}+2\ln x +3-\frac{\pi^2}{3}
\Bigg)H^{(0)}(x,Q^2,k_T)\;.\label{pht2}
\end{eqnarray}
Employing the factorization scale $\mu_{\rm f}=Q$ and the
asymptotic model of the pion wave function, the same as in the LO
analysis in $k_T$ factorization theorem \cite{KR96}, the NLO
corrections are found to be only 5\%. That is, the NLO corrections
are not expected to affect much the LO results for $\pi\gamma^*\to
\gamma$. Our conclusion is drawn under the specific factorization
scheme with $\nu=\nu_u=1$. It requires an examination whether NLO
corrections are also negligible under the same scheme in other
exclusive processes containing pions, such as the pion form factor
involved in $\pi\gamma^*\to\pi$.

\section{GAUGE INVARIANCE}

\begin{figure}[tb]
\begin{center}
\includegraphics[scale=0.8]{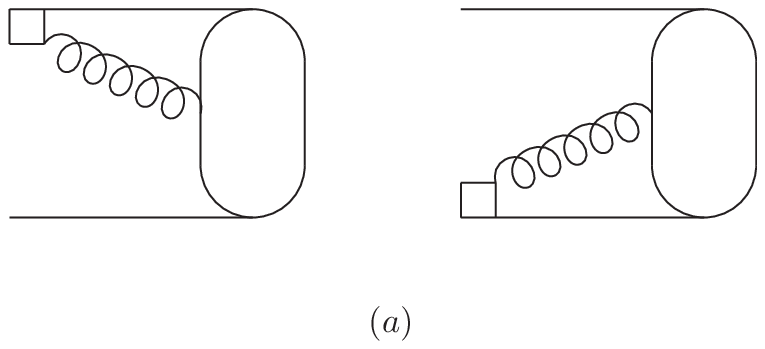}
\vskip 0.5cm
\includegraphics[scale=0.8]{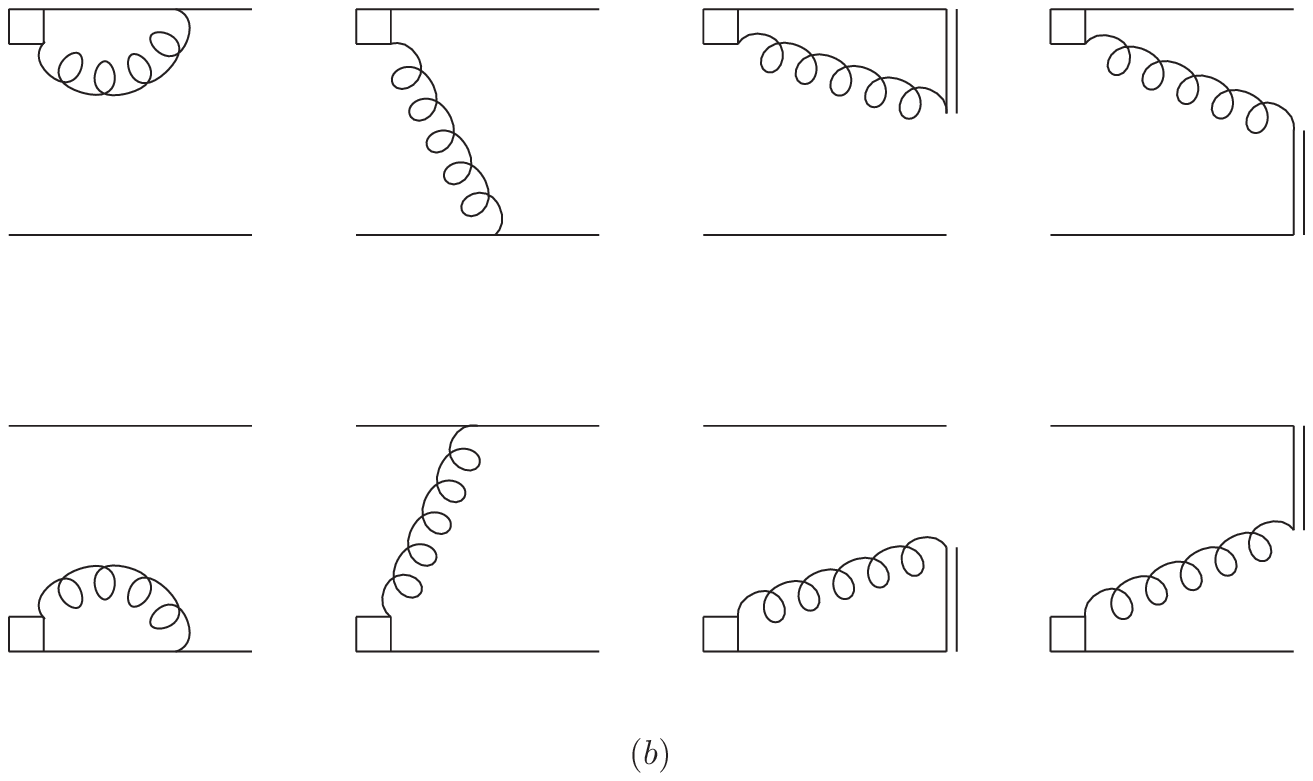}
\caption{(a) Diagrams for $\lambda dG^{(N+1)}/d\lambda$, where the
bubbles represent $G^{(N)}$, and the squares contain the special
vertex $v_{\alpha}$. (b) Diagrams for $\lambda
d\Phi^{(1)}/d\lambda$.}\label{fig4}
\end{center}
\end{figure}

In this section we prove the gauge invariance of the
$k_T$-dependent hard kernel for the pion transition form factor by
induction. We first show that the $k_T$ factorization constructed
in the Feynman gauge \cite{NL2,L1} holds in an arbitrary covariant
gauge $\partial\cdot A=0$ with the gauge parameter $\lambda$, in
which the gluon propagator is given by $(-i/l^2)N^{\mu\nu}(l)$
with the tensor
\begin{equation}
N^{\mu\nu}(l)=g^{\mu\nu}-\left(1-{\lambda}\right) \frac{l^\mu
l^\nu}{l^2}\;. \label{gp}
\end{equation}
It has been argued that the replacement
\begin{equation}
g^{\mu\nu}\to \frac{n_-^\mu l^\nu}{n_-\cdot l}=g^{\mu\alpha}
\frac{n_{-\alpha} l^\nu}{n_-\cdot l}\;,
\end{equation}
for a collinear gluon propagator in the Feynman gauge extracts
collinear divergences correctly \cite{NL2,L1}. In the arbitrary
covariant gauge we just need to modify the above replacement into
\begin{equation}
N^{\mu\nu}(l)\to \frac{n_-^\mu l^\nu}{n_-\cdot
l}-\left(1-{\lambda}\right) \frac{l^\mu
l^\nu}{l^2}=N^{\mu\alpha}(l)\frac{n_{-\alpha} l^\nu}{n_-\cdot
l}\;, \label{gpr}
\end{equation}
and then the procedures for deriving factorization theorem in
\cite{NL2,L1} follow: the Ward identity is applied to all the
contractions of $l^\nu$, leading to the factorization of the
collinear gluon. The factor $n_{-\alpha}/n_-\cdot l$ explains how
the Wilson lines are generated in factorizing hadron wave
functions. For more details, refer to \cite{NL2,L1}.

The $k_T$ dependence in a hard kernel implies that the partons
entering the quark diagrams and the effective diagrams for the
pion wave function are off-shell by $k_T^2$. The LO hard kernel
$H^{(0)}(x,Q^2,k_T)$ in Eq.~(\ref{h0p}), which does not contain a
gluon, is independent of the gauge parameter $\lambda$. Beyond LO,
the gauge invariance of a hard kernel is a consequence of the
gauge-dependence cancellation between the above two sets of
diagrams. Assuming that the hard kernels defined by
\begin{eqnarray}
H^{(j)}(x,Q^2,k_T)&=&G^{(j)}(x,Q^2,k_T) -\sum_{i=1}^{j}\int dx'
d^2k'_T \Phi^{(i)}(x,k_T;x',k'_T)H^{(j-i)}(x',Q^2,k'_T)\;,
\label{gan}
\end{eqnarray}
are gauge-invariant for $j=1,2,\cdots N$, we shall prove the gauge
invariance of the $O(\alpha_s^{N+1})$ hard kernel
\begin{eqnarray}
H^{(N+1)}(x,Q^2,k_T)&=&G^{(N+1)}(x,Q^2,k_T) -\sum_{i=1}^{N+1}\int
dx' d^2k'_T \Phi^{(i)}(x,k_T;x',k'_T)H^{(N+1-i)}(x',Q^2,k'_T)\;,
\label{gann}
\end{eqnarray}
using the method proposed in \cite{CLY}. Note that the external
quark spinors in the nonlocal matrix element in Eq.~(\ref{de1})
have absorbed half of the self-energy corrections. Another half
goes into the higher-order wave functions, giving the coefficients
$1/2$ in Eqs.~(\ref{pka}) and (\ref{pkb}). The same explanation
applies to the appearance of $1/2$ in Eqs.~(\ref{pga}) and
(\ref{pgb}) for the $O(\alpha_s)$ quark diagrams. To discuss the
gauge dependence, we consider the full self-energy corrections to
the quark diagrams $G$ and to the effective diagrams $\Phi$.

Applying the differential operator $\lambda d/d\lambda$ to
$H^{(N+1)}$, it acts only on the gluon propagators in $G^{(N+1)}$
and $\Phi^{(i)}$ on the right-hand side of Eq.~(\ref{gann}),
leading to
\begin{eqnarray}
{\lambda}\frac{d}{d\lambda}N^{\mu\nu}= {\lambda}\frac{l^\mu
l^\nu}{l^2} =v_\alpha(l^\mu N^{\alpha\nu}+N^{\mu\alpha}l^\nu)\;,
\label{digp}
\end{eqnarray}
with the special vertex $v_\alpha=l_\alpha/(2l^2)$. The
derivatives $\lambda dH^{(N+1-i)}/d\lambda$ vanish due to the
gauge-invariant assumption associated with Eq.~(\ref{gan}). The
loop momentum $l^\mu$ ($l^\nu$) in Eq.~(\ref{digp}) contracts with
vertices in the diagrams of $G^{(N+1)}$ and $\Phi^{(i)}$, which
are then replaced by the special vertex $v_\alpha$. Summing all
the quark diagrams with various differentiated gluons and
employing the Ward identity, only those, in which the special
vertex is located at the outer ends of the valence quark lines,
are left \cite{L1,CLY} as shown in Fig.~\ref{fig4}(a). These
diagrams come from the second terms in the following Ward
identities associated with the quark and the anti-quark,
respectively,
\begin{eqnarray}\label{eoq}
& &\frac{i(\not \bar k+\not l)}{(\bar k+l)^2}(-i\not l)\not
P_1\gamma_5 =\not P_1\gamma_5-\frac{\not \bar k+\not l}{(\bar
k+l)^2}\not \bar k \not P_1\gamma_5\;,
\nonumber\\
& &\not P_1\gamma_5(-i\not l)\frac{i(\not l-\not k)}{(l- k)^2}
=\not P_1\gamma_5+\not P_1\gamma_5\not k\frac{\not l-\not
k}{(l-k)^2}\;,
\end{eqnarray}
where $\not P_1\gamma_5$ is the leading spin structure appearing
in the expressions for the quark diagrams. We have the similar
Ward identities for the effective diagrams with $\not n_+\gamma_5$
being substituted for $\not P_1\gamma_5$. The first term is
cancelled by one of the two terms from the contraction of $l$ with
the adjacent vertex. If all the external quarks were on mass
shell, ie., $k_T=0$, the second terms also vanish due to $\not
\bar k \not P_1=\not P_1\not k=0$, implying $\lambda
dG^{(N+1)}/d\lambda=0$ and $\lambda d\Phi^{(i)}/d\lambda=0$. That
is, the quark diagrams from full QCD and the effective diagrams
for the wave function with on-shell partons are gauge-invariant.

For the differentiated quark diagrams $G^{(N+1)}$ in
Fig.~\ref{fig4}(a), the gluon emitting from the special vertex
attaches all the lines inside $G^{(N)}$. Adopting Eq.~(\ref{gpr})
and the procedures in \cite{L1}, $\lambda dG^{(N+1)}/d\lambda$ is
factorized into the convolution of $G^{(N)}$ with the
differentiated $\Phi^{(1)}$ at leading power in $1/Q$:
\begin{eqnarray}
{\lambda}\frac{d}{d\lambda}G^{(N+1)}(x,Q^2,k_T)=\int dx' d^2k'_T
{\lambda}\frac{d}{d\lambda}\Phi^{(1)}(x,k_T;x',k'_T)G^{(N)}(x',Q^2,k'_T)\;.
\label{dg}
\end{eqnarray}
For illustration, we display the effective diagrams for $\lambda
d\Phi^{(1)}/d\lambda$ in Fig.~\ref{fig4}(b) explicitly. We repeat
the above steps for the differentiated wave function
$\Phi^{(i)}(x,k_T;x',k'_T)$, and obtain
\begin{eqnarray}
{\lambda}\frac{d}{d\lambda}\Phi^{(i)}(x,k_T,x'',k''_T)=\int dx'
d^2k'_T{\lambda}\frac{d}{d\lambda}\Phi^{(1)}(x,k_T;x',k'_T)
\Phi^{(i-1)}(x',k'_T;x'',k''_T) \;.\label{dph}
\end{eqnarray}
Combining Eqs.~(\ref{dg}) and (\ref{dph}), the differentiation of
the $O(\alpha_s^{N+1})$ hard kernel gives
\begin{eqnarray}
{\lambda}\frac{d}{d\lambda}H^{(N+1)}(x,Q^2,k_T)&=&\int dx' d^2k'_T
{\lambda}\frac{d}{d\lambda}\Phi^{(1)}(x,k_T;x',k'_T)\nonumber\\
& &\times\left[G^{(N)}(x',Q^2,k'_T) -\sum_{i=0}^{N}\int dx''
d^2k''_T
\Phi^{(i)}(x',k'_T;x'',k''_T)H^{(N-i)}(x'',Q^2,k''_T)\right]\;,
\label{gnn}
\end{eqnarray}
where the term in the square brackets diminishes because of
Eq.~(\ref{gan}) for $j=N$. We then prove the gauge invariance of
the $O(\alpha_s^{N+1})$ hard kernel. The hard kernels and the
resultant predictions from the $k_T$ factorization theorem are
thus gauge-invariant to all orders by induction.

\section{CONCLUSION}

In this paper we have elucidated the framework for the
higher-order calculations in $k_T$ factorization theorem, which is
appropriate for QCD processes dominated by contributions from
small momentum fractions. The point is that partons in both the
quark diagrams from full QCD and the effective diagrams for hadron
wave functions are off mass shell by $k_T^2$. Their difference
gives the gauge-invariant $k_T$-dependent hard kernels, since the
gauge dependences cancel between the two sets of diagrams. The
gauge invariance of the hard kernels for the scattering process
$\pi\gamma^*\to\gamma$ in $k_T$ factorization theorem has been
proven to all orders by induction. The proof can be easily
generalized to other processes. We have explained that the
light-cone divergences in a naive definition of $k_T$-dependent
hadron wave functions are regularized by rotating the Wilson lines
away from the light cone. This procedure introduces a
regularization-scheme dependence, which, however, can be regarded
as part of the factorization-scheme dependence, and minimized by
adhering to a fixed prescription for deriving hard kernels. The
gauge invariance of a hard kernel and the removal of the
light-cone singularities are the two essential ingredients for
making physical predictions from $k_T$ factorization theorem.

We have calculated the NLO $k_T$-dependent hard kernel for
$\pi\gamma^*\to\gamma$ in the region with a large momentum
transfer $Q^2$ and a small momentum fraction $x$. We have
demonstrated that the infrared logarithms $\ln k_T^2$, reflecting
the collinear divergences, cancel between the quark diagrams and
the effective diagrams exactly. Hence, there is no need to
introduce the additional nonperturbative soft function in $k_T$
factorization theorem. The quark diagrams generate the double
logarithms $\ln^2(Q^2/k_T^2)$ and $\ln^2 x$ from the loop
correction to the virtual photon vertex. It has been shown that
the former is absorbed into the pion wave function, and the latter
into the jet function, confirming the observations made in our
previous works \cite{LY1,UL}. Note that the factorization of the
jet function does not alter the renormalization-group behavior of
the hard kernel. Eventually, the NLO corrections in
$\pi\gamma^*\to\gamma$ amount only to 5\% under a specific
factorization scheme with the factorization scale set to the
momentum transfer. NLO corrections under the same factorization
scheme in other processes, such as the pion form factor and
heavy-to-light transition form factors, will be examined
elsewhere. At last, we mention that the off-light-cone effects
from $y_T\not=0$ have been found to be sizable in some
heavy-to-light correlators based on a QCD-sum-rule analysis
recently \cite{LMS07}.

\vskip 0.3cm We thank X. Ji, J.P. Ma, Y.L. Shen, I. Stewart, Q.
Wang, and H. Zou for useful discussions. The work was supported
in part by the National Science Council of R.O.C. under Grant No.
NSC-95-2112-M-050-MY3, by the National Center for Theoretical
Sciences of R.O.C., and by UGC (India) research fellowship. SN
thanks Institute of Physics, Academia Sinica and Department of
Physics, National Taiwan University for their hospitality during
his visit for this project.


\begin{thebibliography}{99}

\bibitem{CCH} S. Catani, M. Ciafaloni and F. Hautmann, Phys. Lett.
B {\bf 242}, 97 (1990); Nucl. Phys. {\bf B366}, 135 (1991).
\bibitem{CE} J.C. Collins and R.K. Ellis, Nucl. Phys. {\bf B360}, 3 (1991).
\bibitem{LRS} E.M. Levin, M.G. Ryskin, Yu.M. Shabelskii,
and A.G. Shuvaev, Sov. J. Nucl. Phys. {\bf 53}, 657 (1991).
\bibitem{BS} J. Botts and G. Sterman, Nucl. Phys. {\bf B325}, 62 (1989).
\bibitem{LS} H-n. Li and G. Sterman, Nucl. Phys. {\bf B381}, 129 (1992).
\bibitem{HS} T. Huang and Q.X. Shen, Z. Phys. C {\bf 50}, 139 (1991);
J.P. Ralston and B. Pire, Phys. Rev. Lett. {\bf 65}, 2343 (1990);
R. Jakob and P. Kroll, Phys. Lett. B {\bf 315}, 463 (1993); B {\bf
319}, 545 (1993)(E).
\bibitem{NL2} M. Nagashima and H-n. Li, Phys. Rev. D {\bf 67},
034001 (2003).
\bibitem{LY1} H-n. Li and H.L. Yu, Phys. Rev. Lett. {\bf 74}, 4388 (1995);
Phys. Lett. B {\bf 353}, 301 (1995); Phys. Rev. D {\bf 53}, 2480
(1996).
\bibitem{CL} C.H. Chang and H-n. Li, Phys. Rev. D {\bf 55}, 5577 (1997).
\bibitem{YL} T.W. Yeh and H-n. Li, Phys. Rev. D {\bf 56}, 1615 (1997).
\bibitem{KLS} Y.Y. Keum, H-n. Li, and A.I. Sanda,
Phys. Lett. B {\bf 504}, 6 (2001); Phys. Rev. D {\bf 63}, 054008
(2001); Y.Y. Keum and H-n. Li, Phys. Rev. {\bf D63}, 074006
(2001).
\bibitem{LUY} C. D. L\"{u}, K. Ukai, and M. Z. Yang, Phys. Rev. D {\bf 63},
074009 (2001).
\bibitem{BL} G.P. Lepage and S.J. Brodsky, Phys. Lett. B {\bf 87},
359 (1979); Phys. Rev. D {\bf 22}, 2157 (1980).
\bibitem{ER} A.V. Efremov and A.V. Radyushkin, Phys. Lett. B {\bf 94},
245 (1980).
\bibitem{CZS} V.L. Chernyak, A.R. Zhitnitsky, and V.G. Serbo,
JETP Lett. {\bf 26}, 594 (1977).
\bibitem{CZ} V.L. Chernyak and A.R. Zhitnitsky,
Sov. J. Nucl. Phys. {\bf 31}, 544 (1980); Phys. Rep. {\bf 112}, 173 (1984).
\bibitem{BBNS}  M. Beneke, G. Buchalla, M. Neubert, and C.T. Sachrajda,
Phys. Rev. Lett. {\bf 83}, 1914 (1999); Nucl. Phys. {\bf B591},
313 (2000).
\bibitem{BPS} C.W. Bauer, S. Fleming, D. Pirjol, and I.W. Stewart, Phys. Rev.
D {\bf 63}, 114020 (2001).
\bibitem{Co03} J.C. Collins, Acta. Phys. Polon. B {\bf 34}, 3103 (2003).
\bibitem{LL04} H-n. Li and H.S. Liao, Phys. Rev. D {\bf 70}, 074030 (2004).
\bibitem{MW} J.P. Ma and Q. Wang, JHEP {\bf 0601}, 067 (2006);
Phys. Lett. B {\bf 642}, 232 (2006).
\bibitem{Neu03} B.O. Lange and M. Neubert, Phys. Rev. Lett. {\bf 91},
102001 (2003).
\bibitem{BIK} V.M. Braun, D.Yu. Ivanov, and G.P. Korchemsky,
Phys. Rev. D {\bf 69}, 034014 (2004).
\bibitem{CS} J.C. Collins and D.E. Soper, Nucl. Phys. {\bf B193}, 381
(1981).
\bibitem{MR} I.V. Musatov and A.V. Radyushkin, Phys. Rev. D {\bf 56},
2713 (1997).
\bibitem{TLS} T. Kurimoto, H-n. Li, and A.I. Sanda,
Phys. Rev D {\bf 65}, 014007 (2002).
\bibitem{WY} Z.T. Wei and M.Z. Yang, Nucl. Phys. {\bf B642}, 263 (2002).
\bibitem{LRev} H-n. Li, Prog. Part. Nucl. Phys. {\bf 51}, 85 (2003);
Czech. J. Phys. {\bf 53}, 657 (2003).
\bibitem{KPY} G.P. Korchemsky, D. Pirjol, and T.M. Yan,
Phys. Rev. D {\bf 61}, 114510  (2000).
\bibitem{MW0607} J.P. Ma and Q. Wang, Phys. Rev. D {\bf 75}, 014014 (2007)
and references therein.
\bibitem{NLO} F. del Aguila and M.K. Chase, Bucl. Phys. {\bf
B193}, 517 (1981); E. Braaten, Phys. Rev. D {\bf 28}, 524 (1983);
E.P. Kadantseva, S.V. Mikhailov, and A.V. Radyushkin, Yad. Fiz.
{\bf 44}, 507 (1986) [Sov. J. Nucl. Phys. {\bf 44}, 326 (1986)].
\bibitem{NNLO}  B. Melic, D. Muller, and K. Passek-Kumericki,
Phys. Rev. D {\bf 68}, 014013 (2003).
\bibitem{UL} H-n. Li, Phys. Rev. D {\bf 66}, 094010 (2002);
K. Ukai and H-n. Li, Phys. Lett. B {\bf 555}, 197 (2003).
\bibitem{CKL} C.H. Chen, Y.Y. Keum, and H-n. Li,
Phys. Rev. D {\bf 64}, 112002 (2001).
\bibitem{L1} H-n. Li, Phys. Rev. D {\bf 64}, 014019 (2001); M. Nagashima
and H-n. Li, Eur. Phys. J. C {\bf 40}, 395 (2005).
\bibitem{BJY} X. Ji, and F. Yuan, Phys. Lett. B {\bf 543}, 66 (2002);
A.V. Belitsky, X. Ji, and F. Yuan, Nucl. Phys. {\bf B656}, 165
(2003).
\bibitem{Li98} H-n. Li, hep-ph/9803202.
\bibitem{WS79} W. Siegel, Phys. Lett. B {\bf 84}, 193 (1979).
\bibitem{ASY} R. Ahkoury, G. Sterman, and Y.P. Yao,
Phys. Rev. D {\bf 50}, 358 (1994).
\bibitem{Li96} H-n. Li, Phys. Rev. D {\bf 55}, 105 (1997).
\bibitem{LL} H-n. Li, Phys. Lett. B {\bf 405}, 347 (1997);
hep-ph/9703328; H-n. Li and J.L. Lim, Eur. Phys. J. C {\bf 10},
319 (1999).
\bibitem{KR96} P. Kroll and M. Raulfs, Phys. Lett. B {\bf 387}, 848 (1996).
\bibitem{CLY} H.Y. Cheng, H-n. Li, and K.C. Yang,
Phys. Rev. D {\bf 60}, 094005 (1999).
\bibitem{LMS07} W. Lucha, D. Melikhov, and S. Simula, Phys. Rev. D {\bf 75}, 096002
(2007). 





\end{thebibliography}
\end{document}